\documentclass[times,sort&compress,3p]{elsarticle}
\usepackage[labelfont=bf]{caption}

\usepackage{amsmath,amsfonts,amssymb,amsthm,booktabs,color,epsfig,graphicx,hyperref,url}
\usepackage{dsfont}
\theoremstyle{plain}

\newtheorem{lemma}{Lemma}

\newtheorem{ass}{Assumption}

\theoremstyle{definition}
\newtheorem{definition}{Definition}

\newtheorem{example}{Example}
\newcommand{\bx}{\boldsymbol{x}}
\newcommand{\bxo}{\boldsymbol{x}^{\text{obs}}}
\newcommand{\bXo}{\boldsymbol{X}^{\text{obs}}}
\newcommand{\bXot}{\boldsymbol{X}^{\text{obs}\top}}

\newcommand{\bXm}{\boldsymbol{X}^{\text{miss}}}
\newcommand{\bXmt}{\boldsymbol{X}^{\text{miss}\top}}
\newcommand{\br}{\boldsymbol{r}}
\newcommand{\bX}{\boldsymbol{X}}
\newcommand{\bR}{\boldsymbol{R}}
\newcommand{\bZ}{\boldsymbol{Z}}
\newcommand{\btheta}{\boldsymbol{\theta}}
\newcommand{\btau}{\boldsymbol{\tau}}

\DeclareMathOperator*{\argmax}{arg\,max}

\begin{document}

\begin{frontmatter}

\title{Clustering Data with Non-Ignorable Missingness using Semi-Parametric Mixture Models}
\author[label1]{Marie Du Roy de Chaumaray}
\author[label1]{Matthieu Marbac\corref{mycorrespondingauthor}}

\address[label1]{Univ. Rennes, Ensai, CNRS, CREST - UMR 9194, F-35000 Rennes, France}

\cortext[mycorrespondingauthor]{Corresponding author. Email address: \url{matthieu.marbac-lourdelle@ensai.fr}}

\begin{abstract}
We are concerned in clustering continuous data sets subject to non-ignorable missingness.
 We perform clustering with a specific semi-parametric mixture, under the assumption  of conditional independence given the component. 
The mixture model is used for clustering and not for estimating the density of the full variables (observed and unobserved), thus we do not need other assumptions on the component distribution or to specify the missingness mechanism. 
Estimation is performed by maximizing an extension of smoothed likelihood allowing missingness. This optimization is achieved by a Majorization-Minorization algorithm.   We illustrate the relevance of our approach by numerical experiments on simulated and benchmark data. Under mild assumptions, we show the identifiability of the model defining the distribution of the observed data and the monotonicity of the algorithm. 
We also propose an extension of this new method to the case of mixed-type data that we illustrate on a real data set. The proposed method is implemented in the R package \texttt{MNARclust} available on CRAN.\end{abstract}


\end{frontmatter}

\section{Introduction}
Clustering is a useful tool to analyze large data sets because it aims to group the subjects into few homogeneous subpopulations. In this context, mixture models permit to achieve the clustering purpose \citep{McL00,fruhwirth2019handbook} since they model the distribution of the observed data. Despite the fact that the data sets often contain missing values, like in social surveys, there are only few clustering approaches that consider missingness.  Thus,  statistical analysis are generally performed on a complete data  set where missing values have been either removed or imputed. Removing subjects having missing values leads to severe bias and/or losses of efficiency \citep{molenberghsJRSSB2008}. Single imputation of missing values \citep{van2018flexible} suffers from a lack of consistency because imputations are generally performed with a model different to the model used to cluster and do not permit to consider the variability of the data.

When the missingness mechanism is \emph{ignorable} (\emph{i.e.,} the mechanism is \emph{Missing at Random} and the property of distinctness is satisfied; \cite{molenberghs_2014,little2019statistical}), then the distribution of the variables can be estimated by modeling the missingness mechanism. Thus, if the parameter of the distribution of the variables is the quantity of interest, likelihood-based methods \citep{schafer_1997} or multiple imputations \citep{van2018flexible} can be used for estimation. Note that, in this paper, the quantity of interest is not the distribution of the variables but the conditional probability of the cluster memberships given the observed variables.

The case where the mechanism is not ignorable (\emph{e.g.,} \emph{missing not at random} (MNAR) mechanism, where the missingness depends on the missing values even conditionally  on the observed covariates) happens frequently in practice (\emph{e.g.,} higher-income respondents may decline to report income data). In such cases, the joint distribution of the variables and the indicators of responses has to be considered. Thus, weighting methods \citep{rotnitzky_1997,tsiatis_2007} can be used if the target is the inference of the distribution of the variables. However, these methods are not really suitable for clustering because they would classify only the subjects with no missingness. Alternatively, multiple imputations could be considered, but because many samples would be generated, it is not easy to consider the aim of the clustering. Thus, we focus on the likelihood-based method. Note that generally, assumptions should be made (\emph{e.g.,} parametric assumption) on the joint distribution of the  variables and the indicators of responses to obtain the identifiability of the model but the distribution of the mechanism cannot be tested on the observed data \citep{molenberghsJRSSB2008}. Identifiability of the parameter of interest is crucial for consistency of the procedure.


Two clustering approaches allow data subject to non-ignorable mechanism  to be analyzed. Thus, \citet{chiAmer2016} introduces the $K$-POD  algorithm that extends the $K$-means to the case of missing data even if the missing mechanism is unknown. However, this approach suffers from the standard drawbacks of the $K$-means algorithm (\emph{i.e.,} assumptions of spherical clusters and equals proportions of the clusters). Alternatively, using a \emph{selection model} approach (see \citet{little1993pattern} and the definition in Section~\ref{sec:model}), \citet{miaoJASA2016}  proposed specific Gaussian mixtures and $t$-mixtures to cluster continuous data under a non-ignorable mechanism. For such an approach, the missingness mechanism must be specified; probit and logit distributions are generally used. However, this approach produces strong bias if the parametric assumptions, made either on the covariate distribution or on the missingness mechanism, are violated.

In this paper, clustering is performed via a mixture model that uses a \emph{pattern-mixture model} approach (see \citet{little1993pattern} and the definition in Section~\ref{sec:model}) with non-parametric distributions. Thus, no assumptions are made on the data distribution or on the missingness mechanism except that the variables are independent within components. Note that this assumption is quite standard for semi-parametric mixtures \citep{hallAOS2003,kasaharaJRSSB2014,chauveauSurveys2015,zhengJASA2019,kwon2019estimation}.
 Moreover, this is an implicit assumption made by geometrical clustering (\emph{e.g.,} $K$-means or $K$-POD) when a diagonal metric is used to compute the distances between observations. Finally, note that the parametric mixture of \citet{miaoJASA2016} is introduced to cluster univariate data and that its extension to the case of multivariate data is not trivial without the assumption of independence within components. Despite that this assumption is relevant in many situations, especially if the number of variables is large with respect to the number of observations \citep{hand2001idiot,webb2005not,stephens2018naive}, it can induce a bias when violated. Thus, we discuss in Section~\ref{sec:concl} how this assumption can be relaxed.  For each mixture component, we estimate, for each variable, its probability to be observed together with its conditional distribution given that the variable is observed. 
 We emphasize that our concern is clustering and not imputation or density estimation. Indeed,  the approach permits to estimate the conditional probability of the cluster memberships given the observed values. Note that, like in any approach developed for a non-ignorable mechanism,   the distribution of the variables within component cannot be estimated by our procedure, without additionnal assumptions.
 
  Estimation of the semi-parametric mixture can be done by maximizing the smoothed likelihood \citep{levineBiometrika2011}. In this paper, we extend the concept of smoothed likelihood to mixed-type data. Indeed, the model includes continuous (the covariates) as well as binary (the indicators of the missigness) variables. In our extension, only the distribution of the continuous variables are smoothed. Thus, the smoothed likelihood can be maximized by a Majorization-Minorization (MM) algorithm \citep{hunter2004tutorial}.

The paper is organized as follows. Section~\ref{sec:model} introduces the semi-parametric mixture used for clustering data with non-ignorable missingness  and a definition of ignorability for clustering that can be interpreted as an extension of the ignorability for a part of the parameters introduced in \citet{little2017conditions}. Section~\ref{sec:estim} presents the MM algorithm used for estimation. Section~\ref{sec:exp} illustrates the relevance of the approach on numerical experiments. Section~\ref{sec:concl} gives a conclusion and further discussion on the assumptions. Proofs of the theoretical results (model identifiability and monotonicity of the MM algorithm) are presented in \ref{appendix:proof}. Details on the extention of the approach to case of clustering mixed-type data are given in \ref{app:extension}. Finally, details on the simulations and on the real data analysis are given in \ref{appendix:exp} and \ref{app:echo}. 

\section{Mixture for non-ignorable missingness}\label{sec:model}
\subsection{The data}
The observed sample is composed of $n$ independent and identically distributed subjects arisen form $K$ homogeneous subpopulations. Each subject is described by $d$ continuous variables and some realizations of these variables may be unobserved. The missingness mechanism is allowed to be non-ignorable. Thus, the probability, for a variable,  to be not observed is allowed to depend on the values of the variable itself and the subpopulation membership.

Each subject $i$ is described by a vector of three variables $(\bX_i^\top, \bR_i^\top, \bZ_i^\top)^\top$ where $\bX_i=(X_{i1},\ldots,$ $X_{id})^\top\in\mathbb{R}^d$ is a set of continuous variables, $\bR_i=(R_{i1},\ldots,R_{id})^\top\in\{0,1\}^d$ indicates whether $X_{ij}$ is observed ($R_{ij}=1$) and $\bZ_i=(Z_{i1},\ldots,Z_{iK})^\top$ indicates the subpopulation of subject $i$ ($Z_{ik}=1$ if subject $i$ belongs to subpopulation $k$ and otherwise $Z_{ik}=0$). Each subject belongs to one subpopulation such that $\sum_{k=1}^K Z_{ik}=1$. The realizations of $\bZ_i$ are unobserved and a part of the realizations of $\bX_i$ can be unobserved too. Therefore, the observed variables for subject $i$ are $(\bXot_i,\bR_i^\top)^\top$ where $\bXo_i$ is composed by the elements of $\bX_i$ such that $R_{ij}=1$ and the unobserved variables for subject $i$ are $(\bXmt_i,\bZ_i^\top)^\top$ where $\bXm_i$ is composed by the elements of $\bX_i$ for which $R_{ij}=0$.

\subsection{General mixture model}
We use mixture models for the purpose of clustering and not for density estimation. 
Clustering aims to estimate the subpopulation memberships given the observed variables (\emph{i.e.,} the realization of $\bZ_i$ given $(\bXot_i,\bR_i^\top)^\top$) without any assumption on the missingness mechanism (\emph{i.e.,} no assumption on the conditional distribution of $\bR_i\mid \bX_i,\bZ_i$). The probability distribution function (pdf) of $(\bX_i^\top,\bR_i^\top)^\top$ for subpopulation $k$ (\emph{i.e.,} $Z_{ik}=1$) is denoted by $g_k(\cdot)$. Thus, the pdf $(\bX_i^\top,\bR_i^\top)^\top$ is defined by the pdf of a $K$-component mixture
\begin{equation} \label{eq:mixture1}
g(\bx_i,\br_i) = \sum_{k=1}^K \pi_k g_{k}(\bx_i,\br_i),
\end{equation}
where  $\pi_k>0$, $\sum_{k=1}^K\pi_k=1$ and $g_k(\cdot)$ is pdf of component $k$.
From \eqref{eq:mixture1},  the distribution of the observed values can be defined by two approaches \citep{molenberghs_2014,little2019statistical}: the \emph{selection model} and the \emph{pattern-mixture model}. The approach named \emph{selection model} defines the joint distribution of $(\bX_i^\top,\bR_i^\top)^\top\mid \bZ_i$ as the product between the distribution of $\bX_i\mid \bZ_i$ and the  distribution of $\bR_i\mid \bZ_i,\bX_i$. 

This approach is natural and has been used for a while. When the mechanism is ignorable, it permits an estimation of the marginal distribution of $\bX_i$ without considering the distribution of the mechanism. However, when the mechanism is non-ignorable, it

requires to model the missingness mechanism, \emph{i.e.} the conditional distribution of $\bR_i\mid \bZ_i,\bX_i$. Finally, as it considers the marginal distribution of $\bX_i$, the \emph{selection model} should be used when the aim is to fit the marginal distribution of $\bX_i$. Alternatively, the approach named \emph{pattern-mixture model}  \citep{little1993pattern} defines the joint distribution of $(\bX_i^\top,\bR_i^\top)^\top\mid \bZ_i$ as the product between the distribution of $\bR_i\mid \bZ_i$ and the  distribution of $\bX_i\mid \bZ_i,\bR_i$. Thus, using the \emph{pattern-mixture model}, the pdf of component $k$ is given by
\begin{equation}
g_{k}(\bx_i,\br_i) = g_{k}(\br_i)g_{k}(\bx_i\mid\br_i).\label{eq:mixtobs}
\end{equation}

The pdf of the observed variables under component $k$, denoted by $g_{k}(\bxo_i,\br_i)$, is obtained by integrating the pdf of component $k$ over the missing variables $\bXm_i$, which leads to 
\begin{equation} \label{eq:compo}
g_{k}(\bxo_i,\br_i)  =  g_{k}(\br_i)g_{k}(\bxo_i\mid\br_i). 
\end{equation}
Note that \eqref{eq:compo} takes into account the missingness mechanism as it involves the whole vector $\br_i$. Thus the missing values impact the clustering. 

 To estimate the marginal density of $\bX_i$ from \eqref{eq:compo}, assumptions should be made on the conditional distribution $\bXo_i$ given $\bR_i$ (because the realizations under some distributions are never observed, \emph{e.g.,} $\br_i=0$).  However, we recall that we focus on the target of clustering that consists in assessing the posterior probabilities of the classification given the observed values using 
\begin{equation}
\mathbb{P}(Z_{ik}=1\mid \bxo_i, \br_i) = \frac{\pi_k g_{k}(\bxo_i,\br_i) }{\sum_{\ell=1}^K \pi_\ell g_{\ell}(\bxo_i,\br_i) }. \label{probapost}
\end{equation}
For clustering, the approach named \emph{pattern-mixture model} should be preferred because it turns out to be more general as it does not require the missingness mechanism to be specified and allows this mechanism to be non-ignorable. 
Note that, however, this approach does not permit to estimate the marginal distribution of $\bX_i\mid \bZ_i$ without adding assumptions on the missing mechanism. Thus, the proposed approach can be used for clustering but not for density estimation.  We now discuss the definition of non-ignorable missingness mechanism for clustering.

\subsection{Weak and strong ignorability for clustering}
In a likelihood-based estimation, the missingness mechanism is said to be ignorable for likelihood inference if the missing data are missing at random and if the distinctness property is satisfied by the parameters (see Definition 6.4 in \citet{little2019statistical}). These conditions ensure that it is appropriate to ignore the missingness mechanism, especially for parameter estimation. These conditions have been extended when only a subset of the parameters are of interest \citep{little2017conditions}. Thus, despite that the missingness mechanism is MNAR, these conditions ensure that a subset of the parameters can be consistently estimated by ignoring the missingness mechanism. In clustering, the quantities of interest are the partition and, sometimes, the posterior probabilities of classification. Thus, we introduce the notion of weakly and strongly ignorable mechanisms for clustering that allows the mechanism to be neglected for estimating the partition and the posterior probabilities of classification.

\begin{definition}
Let $f_k(\bxo_i)$ be the marginal pdf of the observed variables under component $k$. The missingness mechanism is said to be strongly ignorable for clustering if 
$$
\forall \bxo_i,\; \frac{\pi_k g_{k}(\bxo_i,\br_i) }{\sum_{\ell=1}^K \pi_\ell g_{\ell}(\bxo_i,\br_i) }=\frac{\pi_k f_{k}(\bxo_i) }{\sum_{\ell=1}^K \pi_\ell f_{\ell}(\bxo_i) }.
$$
The missingness mechanism is said to be weakly ignorable for clustering if
$$
\forall \bxo_i,\; \zeta(\bxo_i) = \eta(\bxo_i),
$$
where $$\zeta(\bxo_i)=\argmax_{k=1,\ldots,K} \frac{\pi_k g_{k}(\bxo_i,\br_i) }{\sum_{\ell=1}^K \pi_\ell g_{\ell}(\bxo_i,\br_i) } \text{ and }
\eta(\bxo_i)=\argmax_{k=1,\ldots,K} \frac{\pi_k f_{k}(\bxo_i) }{\sum_{\ell=1}^K \pi_\ell f_{\ell}(\bxo_i) }.$$
\end{definition}
The strong ignorability for clustering implies the weak ignorability for clustering. When the data are MNAR, the weak ignorability can be interpreted as the condition required for a misspecified model (\emph{e.g.,}  the model of the missingness mechanism or the distribution of the mixture components) to provide a consistent estimator of the partition. Note that in clustering, consistency of the estimated partition does not mean a perfect recovery of the partition but that the estimated partition is asymptotically equivalent to the partition obtained by using the rule of the \emph{maximum a posteriori} on the true posterior probabilities of classification. Thus, a misspecified model can provide a consistent estimator of the partition (\emph{e.g.,} if the data arise from a mixture of two univariate Student distributions with the same degrees of freedom, a consistent estimator of the partition can be obtained by considering a mixture of two univariate Gaussian distributions). As the condition on the missingness mechanism to be weakly ignorable for clustering is quite stringent, we need to introduce an approach based on the joint distribution of $(\bX_i^\top,\bR_i^\top)^\top$ which allows to cluster data under a non-ignorable scenario.

\subsection{Semi-parametric mixture for non-ignorable missingness}
A wide range of literature focuses on models assuming that conditionally on knowing the particular subpopulation the subject $i$
came from, its coordinates  $\bX_i$ are independent. Thus, we extend this model for non-ignorable missingness. The couples of variables $(X_{ij},R_{ij})^\top$ are assumed to be conditionally independent given $\bZ_i$. Thus, the distribution of $\bR_i\mid \bZ_i$ is a product of Bernoulli distributions and the conditional density of $\bX_i\mid \bZ_i,\bR_i$ is defined as the product of univariate densities. Thus, from \eqref{eq:mixtobs}, the pdf of component $k$ is also defined by
\begin{equation} \label{eq:melangeobs3}
 g_{k}(\br_i;\btau_k) = \prod_{j=1}^d\tau_{kj}^{r_{ij}}(1-\tau_{kj})^{1-r_{ij}}
 \text{ and }
 g_{k}(\bx_i\mid\br_i)=\prod_{j=1}^d p_{kj}^{r_{ij}}(x_{ij})  q_{kj}^{1-{r_{ij}}}(x_{ij}),
\end{equation}
where $\btau_k=(\tau_{k1},\ldots,\tau_{kd})$, $\tau_{kj}>0$ is the probability that $X_{ij}$ is observed given that subject $i$ belongs to subpopulation $k$, $p_{kj}(\cdot)$ is the conditional density of $X_{ij}$ given $Z_{ik}=1$ and $R_{ij}=1$ and  $q_{kj}(\cdot)$ is the conditional density of $X_{ij}$ given $Z_{ik}=1$ and $R_{ij}=0$. Thus, clustering is achieved by modeling, for each subpopulation, the marginal probability of missingness and the conditional density given that the variable is observed.
Integrating out the unobserved variables $\bXm_i$ (\emph{i.e.,} the elements of vector $\bX_i$ such that $R_{ij}=0$), we have
\begin{equation} \label{eq:melangeobs1}
g(\bxo_i,\br_i;\btheta)=\sum_{k=1}^K \pi_kg_k(\bxo_i,\br_i;\btheta),
\end{equation}
where  the pdf of component $k$ is a specific version of \eqref{eq:compo} defined by 
\begin{equation} \label{eq:melangeobs2}
 g_k(\bxo_i,\br_i;\btheta) = g_{k}(\br_i;\btau_k)\prod_{j=1}^d p_{kj}^{r_{ij}}(x_{ij}) ,
\end{equation}
where $\btheta$ groups all the finite parameters ($\pi_k$ and $\btau_{k}$) and all the infinite parameters $p_{kj}(\cdot)$. Note, cluster analysis does not require to  estimate $q_{kj}(\cdot)$  because this quantity does not appear in the posterior probabilities of classification given by \eqref{probapost}.
Clustering takes into account the missingness mechanism  because the whole vector $\bR_i$ is considered in \eqref{eq:melangeobs2} and thus in \eqref{probapost}. Thus, missing values impact the posterior probabilities of classification through the parameters $\tau_{kj}$'s used for modeling the binary variables $R_{ij}$. Note that a subject presenting missing value of each variables (\emph{i.e.,} $R_{ij}=0$ for any $j$) has a probability $\pi_{k}g_{k}(\br_i;\btau_k)/\sum_{\ell=1}^K\pi_{\ell}g_{\ell}(\br_i;\btau_\ell)$ to belongs cluster $k$ that is different to the probability obtained under ignorable mechanism (\emph{i.e.,} in this case the probability is $\pi_k$).  Note that the mechanism is strongly ignorable for clustering is also covered by the approach, because such if situation occurs then the conditional distributions of $\bR_i$ given the cluster membership are equal for each cluster (\emph{i.e.,} the vector of probability of responses are equal among components: $\btau_1=\ldots=\btau_K$). Finally, note that the approach allows the mechanism of missingness, for variable $j$, to depend on the cluster membership and/or on the value of the variable itself.  Moreover, model \eqref{eq:melangeobs1}-\eqref{eq:melangeobs2} allows the missing values to have a wide range of influence on the posterior probabilities of classification  as shown by the following example.

\begin{example}[Impact of the missingness mechanism on clustering]
We consider a mixture model of $K$ components such that the distribution of $(\bX_i^\top,\bR_i^\top)^\top$ under component $k$ is defined by
$$
g_k(\bx_i,\br_i)=\prod_{j=1}^d f_j(x_{ij}-\mu_{kj}) \Psi^{r_{ij}}(\gamma_k+\delta_jx_{ij})\left[1-\Psi(\gamma_k+\delta_jx_{ij})\right]^{1-r_{ij}},
$$
where $f_1,\ldots,f_d$ are  known densities and $\Psi$ is a known function defined on $[0,1]$ which represents the missingness mechanism. We show that the distribution of $(\bX_i^\top,\bR_i^\top)^\top$ under component $k$ can be defined  from \eqref{eq:melangeobs3} with
\begin{equation*}
\tau_{kj}=\int  f_j(x_{ij}-\mu_{kj}) \Psi(\gamma_k+\delta_jx_{ij}) dx_{ij},
\end{equation*}
\begin{equation*}
p_{kj}(x_{ij})=\frac{f_j(x_{ij}-\mu_{kj})}{\tau_{kj}} \Psi(\gamma_k+\delta_jx_{ij}) \text{ and }
q_{kj}(x_{ij})=\frac{f_j(x_{ij}-\mu_{kj})}{1-\tau_{kj}}\left[1- \Psi(\gamma_k+\delta_j x_{ij})\right].
\end{equation*}
Clustering is achieved by considering the distribution of the observed values \eqref{eq:melangeobs2}. This framework allows for different situations:
\begin{itemize}
\item The missingness mechanism can depend on the component only (\emph{i.e.,} $\delta_j=0$). If $\gamma_1\neq\ldots\neq\gamma_K$, then the $\tau_{kj}$ are not equals if the $\mu_{kj}$ are not. In this case, $\tau_{kj}=\Psi(\gamma_k)$ and $p_{kj}=q_{kj}=f_j$. Usually when the distributions of a \emph{pattern-mixture model} are equal, then the mechanism is ignorable (see \citet{molenberghs_2014}). However, as the component membership are not observed, the mechanism here is non-ignorable and it can be interpreted as a conditional \emph{MAR} given the component membership.
\item If the mechanism only depends on $j$ (\emph{i.e.,} $\gamma_1=\ldots=\gamma_K$ and $\delta_1\neq\ldots\neq\delta_d$) then $\tau_{kj}$ are different if the $\mu_{kj}$ are. Note that the difference of the $\mu_{kj}$ is required to have different distributions for the mixture components.
\item The clustering is only explained by the mechanism if $\mu_{kj}=0$ and $\gamma_1=\ldots=\gamma_K$.
\item The mechanism is strongly ignorable for clustering (but not for density estimation) if  $\delta_j=1$ and $\mu_{kj}=-\delta_k$ for any $(j,k)$.
\end{itemize}
\end{example}

 Thus, one can consider the case where the partition is only explained by the missing values: \emph{i.e.,} the distribution of $\bX_i$ is the same in each component, but the distribution of $\bR_i\mid\bX_i$ is not. In such case, the probabilities $\tau_{kj}$ are not the same between the components and the distributions of the observed variables per components $p_{kj}$ are generally not the same too. Alternatively, a strongly ignorable missingness mechanism can be considered and this case can be easily detected because it implies that for any $\bR_i\in\{0,1\}^d$, we have for any $(k,\ell)$, $g_k(\br_i;\tau_k)=g_\ell(\br_i;\tau_\ell)$.

With model \eqref{eq:melangeobs1}-\eqref{eq:melangeobs2}, we are able to achieve clustering because the posterior probabilities of classification are available, however, as $q_{kj}(\cdot)$ is not estimated,  we are not able to estimate the distribution of $\bX_i\mid\bZ_i$  or any information on this distribution (\emph{e.g,} the expectation of $X_{ij}\mid\bZ_i$ cannot be computed but only the expectation of  $X_{ij}\mid\bZ_i,R_{ij}=1$).  Avoiding the estimation of $q_{kj}(\cdot)$ is the core of the proposed approach. Indeed, estimating $q_{kj}(\cdot)$ requires information about the missingness mechanism (that is generally unknown) as it can only be achieved by estimating the joint distribution of $(X_{ij},R_{ij})$. \cite{molenberghsJRSSB2008} show that different models used for the distribution of $(X_{ij},R_{ij})$ can lead to the same distribution of the observed variables. Thus, supplementary information is needed for consistently estimating $q_{kj}$. As our approach only considers the marginal probabilities of missingness (for each variable given the component), we avoid the issue of lack of identifiability (see the following lemma) and we are able to  estimate the posterior probabilities of classification (but not the pdf of $\bX_i$ for each component).
 
Sufficient conditions for the model identifiability are stated by Lemma~\ref{lem:iden}. Its proof uses some results on the identifiability of nonparametric mixtures (Theorem~8 of  \citet{allman2009identifiability}) and is postponed in \ref{appendix:proof}. 

\begin{lemma}\label{lem:iden} If $d\geq 3$, $\pi_k>0$ and $\tau_{kj}>0$, and if the densities $p_{kj}$ are linearly independent, then the model defined by \eqref{eq:melangeobs3}-\eqref{eq:melangeobs2} is identifiable, up to label swapping.
\end{lemma}
Note that the assumptions of Lemma~\ref{lem:iden} are not stronger than those of Theorem~8 of  \citet{allman2009identifiability}. 

Indeed, the need to consider at least three variables is explained by the use of the Kruskal's Theorem which is in the core of the results of  \citet{allman2009identifiability}. The assumption of linear independence of the densities is equivalent to the linear independence of the cumulative distribution functions and is not a stringent assumption (see Lemma 17 in \citet{allman2009identifiability}). The fact that identifiability holds up to label swapping is standard in clustering, because the labels of the components of mixture models can be permuted without changing the pdf of the model. 
Finally, note that the assumptions of Lemma~\ref{lem:iden} allow all the $\tau_{kj}$ to be equal to one, corresponding to the case where there is no missingness.
Note that if the data to cluster are univariate or bivariate, the proposed approach cannot be used because model identifiability is not proved. In such cases, alternative models (semi-parametric location-scale model or parametric models) should be considered.

\section{Maximum smoothed likelihood estimate}\label{sec:estim}
\subsection{Smoothed likelihood}
To perform parameter estimation, we extend the approach of  \citet{levineBiometrika2011} that uses the smoothed likelihood to the case of mixed-type variables. Indeed, despite that all the elements of $\bx_i$ are continuous, the vector of indicator of response $\br_i$ is binary, thus the vector of the observed variables $(\bx_i^{\text{obs}\top},\br_i^\top)^\top$ is a vector of mixed-type variables. Note that the smoothing is only performed on the densities  because these quantities are estimated non-parametrically. Thus, smoothing is made on the distributions of $\bxo_i$ for each component and there is no need to smooth the distributions of $\br_i$ for each component because these distribution are just defined as a product of probabilities (see \eqref{eq:melangeobs3}). \\
Let $S$ be the smoothing operator defined by
\begin{equation*}
\mathcal{S}g_{k}(\bxo_i\mid\br_i) =\prod_{j=1}^d \left(\mathcal{S}p_{kj}(x_{ij})\right)^{r_{ij}},
\end{equation*}
and
\begin{equation*}
\mathcal{S}p_{kj}(x_{ij}) = \int_{\Omega_j} \frac{1}{h_j} K\left(\frac{x_{ij} - u}{h_j}\right)p_{kj}(u)du,
\end{equation*}
where $K$ is a kernel function and $h_j>0$ its bandwidth. We consider the non linear smoothing operator defined by
$$
\mathcal{N} g_k(\bxo_i,\br_i;\btheta)=g_k(\br_i;\btau_k)\exp\{ \mathcal{S}\ln g_k(\bxo_i\mid \br_i)\},
$$
where $g_k(\bxo_i\mid \br_i)=\prod_{j=1}^d p_{kj}^{r_{ij}}(x_{ij}) $.\\
The smoothed log-likelihood function is defined by
\begin{equation*}
\ell_n(\btheta) = \sum_{i=1}^n \ln\left( \sum_{k=1}^K \pi_k \mathcal{N} g_k(\bxo_i,\br_i;\btheta)\right).
\end{equation*}
Parameter estimation is performed by maximizing the smoothed likelihood over $\btheta$. This maximization is achieved by a MM algorithm presented in the next section.

\subsection{Majorization-Minimization algorithm}\label{sec:algo}
The maximization on $\btheta$ of the smoothed log-likelihood function is performed via an MM algorithm. This iterative algorithm starts at the initial value of the parameters $\btheta^{[0]}$. At iteration $[r]$, it performs the following two steps
\begin{itemize}
\item Computing the smoothed probabilities of subpopulation memberships $$t_{ik}(\btheta^{[r]}) = \frac{\pi_k^{[r]} \mathcal{N}g_k(\bxo_i,\br_i;\btheta^{[r]})}{\sum_{\ell=1}^K\pi_\ell^{[r]} \mathcal N g_\ell(\bxo_i,\br_i;\btheta^{[r]})}.$$
\item Updating the estimators 
\begin{itemize}
\item Updating of the proportions
$$\pi_k^{[r+1]}= \frac{1}{n} \sum_{i=1}^n t_{ik}(\btheta^{[r]}).$$
\item Updating of the parameters of the missingness mechanism
$$ \tau_{kj}^{[r+1]} = \frac{\sum_{i=1}^n r_{ij} t_{ik}(\btheta^{[r]})}{\sum_{i=1}^n t_{ik}(\btheta^{[r]})}.$$
\item Updating of the conditional distribution
$$
p_{kj}^{[r+1]}(u) = \frac{\sum_{i=1}^n r_{ij} t_{ik}(\btheta^{[r]}) \frac{1}{h_j}K\left(\frac{x_{ij} - u}{h_j}\right)}{\sum_{i=1}^n r_{ij} t_{ik}(\btheta^{[r]})}.
$$
\end{itemize}
\end{itemize}

The monotonicity of the algorithm is stated by Lemma~\ref{lem:monot} whose proof is similar to the proof of Theorem~1 in \citet{levineBiometrika2011} and is postponed to Appendix A. This implies that the algorithm converges to a local optimum of the smoothed log-likelihood, hence different random initializations should be performed. 

\begin{ass}\label{ass:regu}
 For any $1 \leq j \leq d$, any $1 \leq k \leq K$ and any $x_{ij} \in \mathbb{R}$, we suppose that $p_{kj}\in L_1(\mathbb{R})$ and that $\int_{\mathbb{R}} \frac{1}{h_j} K\left(\frac{x_{ij} - u}{h_j}\right) \ln p_{kj}(u)du < +\infty$.
\end{ass}

\begin{lemma}\label{lem:monot}
Let the assumptions of Lemma~\ref{lem:iden} and Assumptions~\ref{ass:regu} hold true. Let $\btheta^{[r]}$ and $\btheta^{[r+1]}$ be the estimators obtained at iterations $[r]$ and $[r+1]$ respectively, we have
$
\ell_n(\btheta^{[r]})\leq \ell_n(\btheta^{[r+1]}).
$
\end{lemma}

%
%
%

\section{Numerical experiments}\label{sec:exp}
This section illustrates the benefits of the proposed method. During all the experiments we use a Gaussian kernel with bandwidths  $h_j=C_jn^{-1/5}$ where $C_j$ is the standard deviation of the observed realizations of variable $j$.
In Section~\ref{subsec:simu}, we compare, on simulated data, our proposed method to standard methods for clustering data with missingness.  In Section~\ref{subsec:appli1}, we illustrate the behavior of the prposed approach on benchmark data where we generate missingness according to different mechanism.
In Section~\ref{subsec:appli2}, we extend of the proposed approach to the case where $\bx_i$ is a vector of mixed-type variables, and we use this extension to analyze a real data.

\subsection{Simulated data}\label{subsec:simu}
In these simulations, different distributions for the components and missingness mechanisms are considered. Moreover, we investigate the influence of four quantities: the rate of missingness, the sample size, the number of variables and the theoretical rate of misclassification. Method comparison is done according to the Adjusted Rand index (ARI;\citep{Hub85}) computed between the true partition and the estimated partition provided by the competing method. Thus, the closer to one the ARI is, the closer the true and the estimated partitions are.

\paragraph{Competing methods}
The proposed method,  implemented in the R package \emph{MNARclust}, is compared to the following three methods:
\begin{itemize}
\item \emph{Ignorable-GMM}: Gaussian mixture assuming that the missingness mechanism is ignorable (implemented in the R package \emph{VarSelLCM}  \citep{marbac2019varsellcm};
\item \emph{K-pod}: $K$-pod approach performed with the function \emph{kpod} of the R package \emph{kpodclustr} \citep{kpod};
\item \emph{NPimputed}: non parametric mixture on the imputed data performed with  the functions \emph{np} and \emph{imputePCA} of the R packages \emph{mixtools} \citep{mixtools} and \emph{missMDA} \citep{missMDA}.
\end{itemize}

\paragraph{Simulation setup}
To compare the different methods of clustering, we generate complete data from  mixture models with three components having unequal proportions ($\pi_1=1/2$ and $\pi_2=\pi_3=1/4$) and  independence between variables within components
such that $$X_{ij}=\delta \sum_{k=1}^3 \lambda_{kj} Z_{ik} + \varepsilon_{ij},$$ where all the $\lambda_{11}=\lambda_{22}=\lambda_{33}=\lambda_{14}=\lambda_{25}=\lambda_{36}=1$ and the other $\lambda_{kj}=0$ and where $\varepsilon_{ij}$ are independent from all the variables and define the distribution within-components (Gaussian, Student with three degrees of freedom, Laplace and Skewed Gaussian with shape equals to three). Then, we add missing values from  four  scenarios:
\begin{itemize}
\item MCAR: $\mathbb{P}(R_{ij}=0\mid X_{ij},\bZ_{i})=(1 + \exp(\gamma))^{-1}$;
\item MNAR-logit-Z: $\mathbb{P}(R_{ij}=0\mid X_{ij},\bZ_{i})=(1 + \exp(\gamma + 2\sum_{k=1}^Kz_{ik}))^{-1}$;
\item MNAR-logit-X: $\mathbb{P}(R_{ij}=0\mid X_{ij},\bZ_{i})=(1 + \exp(\gamma  +  x_{ij}))^{-1}$;
\item MNAR-censoring-X: $\mathbb{P}(R_{ij}=0\mid X_{ij},\bZ_{i})=\mathds{1}_{\{X_{ij}<\gamma\}}$.
\end{itemize}
Thus, the parameters $\delta$ and $\gamma$ allow to set the rates of misclassification and missingness (their values under the different scenarios are given in \ref{appendix:exp}).

\begin{figure}[htp!]
\centering \includegraphics[scale=0.35]{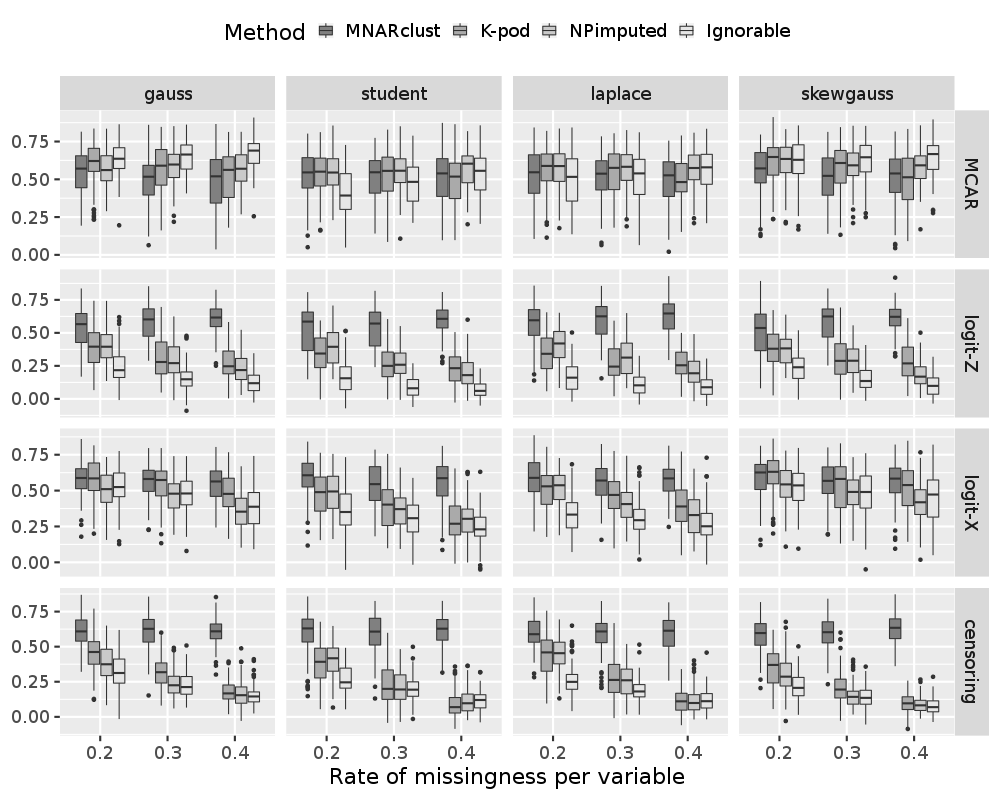}
\caption{Boxplot ARI obtained 100 samples of $n=100$ observations described by $d=6$ variables with a misclassification rate of $10\%$.}\label{fig:simu1}
\end{figure}

\paragraph{Impact of the rate of missingness}
To investigate the impact of the rate of missingness, we consider data sets composed by $n=100$ observations described by $d=6$ variables with a theoretical misclassification of $10\%$ (parameters are given in Table~\ref{tab:paramsim1}). For each scenario, we generated 100 data sets. Figure~\ref{fig:simu1} presents the boxplots of ARI between the true partition and the estimators of the partition given by the methods. 
Overall, the proposed method outperforms the competing methods under non-ignorable mechanims. Indeed, its results are robust to the different distributions of the components, the missingness scenarios and the missingness rates. Moreover, when  the mechanism is ignorable, all the methods obtains good same performances. Note that the parametric approach assuming that the missingness mechanism is ignorable obtains slightly better results, when the distribution within components is  Gaussian or skewed Gaussian. However, this approach obtains poor results when the  missingness mechanism is not ignorable. Under the non-ignorable scenario, the proposed approach obtains the best results. The results of the other methods stays relevant under the logit-X scenario and Gaussian or Skewed-Gaussian components. However, in the other scenarios, they obtain poor results. Finally, note that the larger the missingness rate is, the larger the benefit of the proposed method is. Indeed, the results of the proposed method seems to be not impacted by the missingness rate while the results of the other methods are deteriorated when this rate increasing, under non-ignorable mechanisms.

\begin{figure}[htp!]
\centering \includegraphics[scale=0.35]{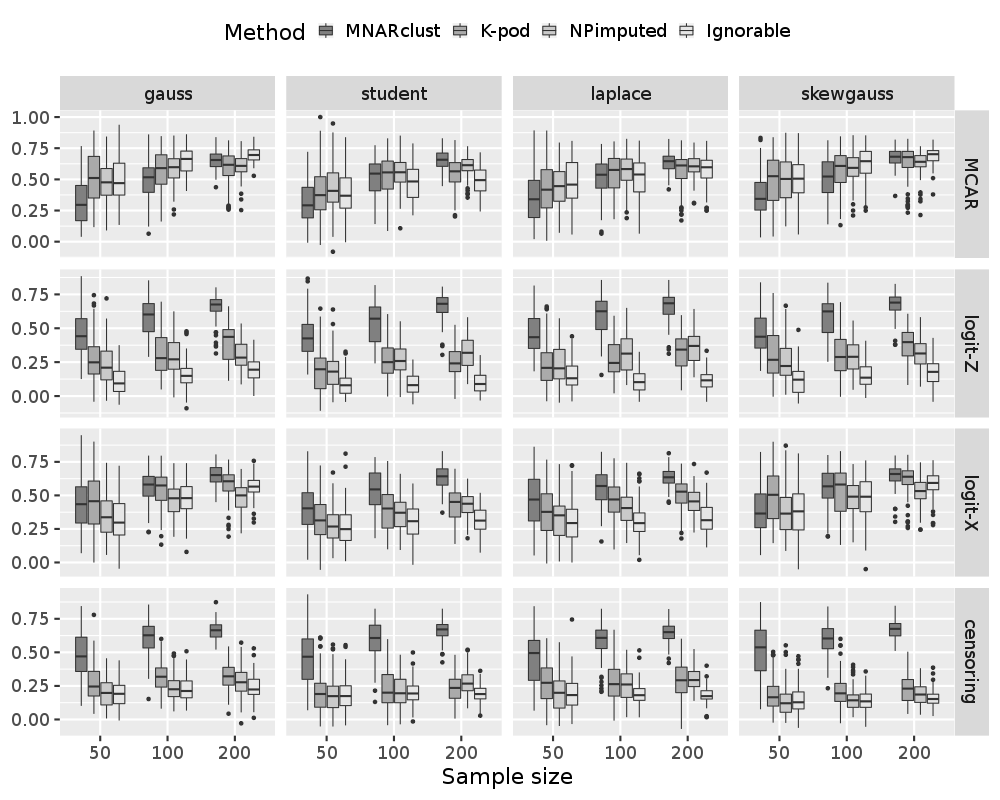}
\caption{Boxplot ARI obtained 100 samples composed of $d=6$ variables having a missing rate of $30\%$ each with a misclassification rate of $10\%$.}\label{fig:simu2}
\end{figure}

\paragraph{Consistency of the estimators}
To illustrate the consistency of the estimators, we consider data sets composed by observations described by $d=6$ variables with a theoretical misclassification of $10\%$ and a theoretical missing rate per variables of $30\%$ (parameters are given in Table~\ref{tab:paramsim1}). For each scenario, we generated 100 data sets. Figure~\ref{fig:simu2} presents the boxplot of the ARI between the true partition and the estimators of the partition given by the methods. 
Again, results show that the method outperforms the competing methods because it is more robust to the distribution of the components and to the missingness scenario. Moreover, despite the accuracy of the partition is improved when the sample size increases, results are satistifying (compared to the results of the parametric methods) even for the small samples. 

\begin{figure}[htp!]
\centering \includegraphics[scale=0.35]{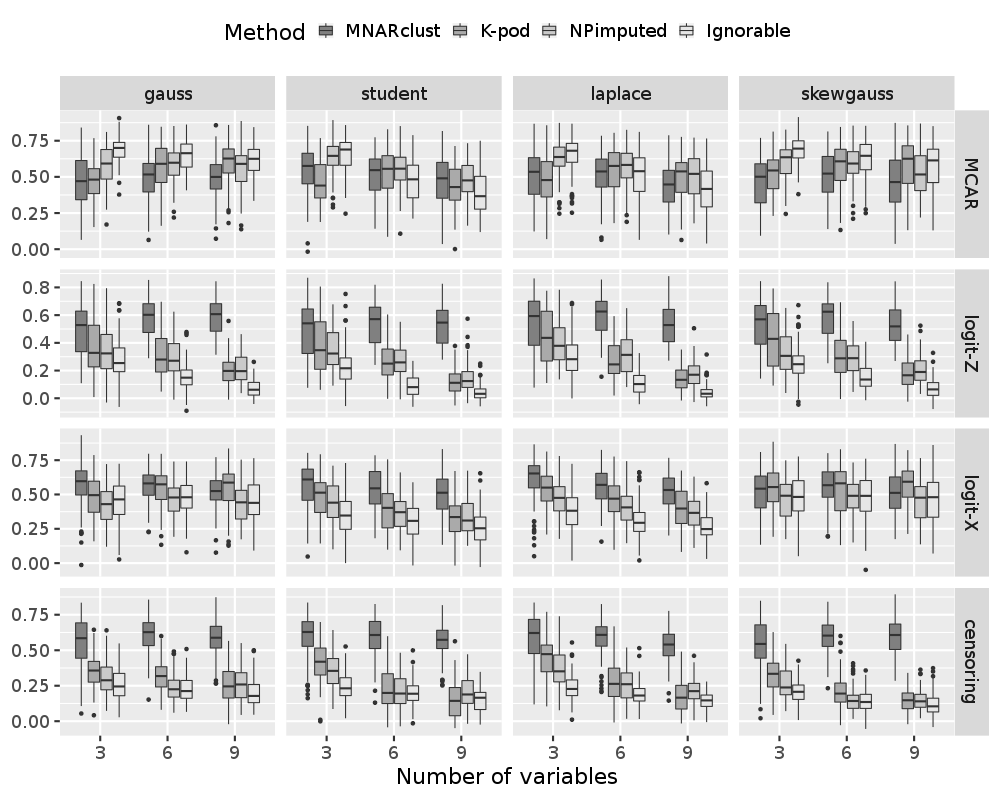}
\caption{Boxplot ARI obtained 100 samples composed of $n=100$ observations having a missing rate of $30\%$ per variable and a misclassification rate of $10\%$.} \label{fig:simu3}
\end{figure}

\paragraph{Impact of the dimension}
To illustrate the impact of the dimension, we consider data sets composed by $n=100$ observations generated with a theoretical misclassification of $10\%$ and a theoretical missing rate per variables of $30\%$ (parameters are given in \ref{tab:paramsim2}). For each scenario, we generated 100 data sets. Figure~\ref{fig:simu3} presents the boxplot of the ARI between the true partition and the estimators of the partition given by the methods. Despite that the proposed method is semi-parametric, results show that it can manage data set with many variables. Indeed, the deterioration of the results of the proposed method when $d$ increases is very weak. This is due to the assumption of conditional independence between the couples $(X_{ij},R_{ij})^\top$ given the components membership. Indeed, this assumption permits to limit the impact of the curse of the dimensionality for the nonparametric estimators.

\begin{figure}[htp]
\centering \includegraphics[scale=0.35]{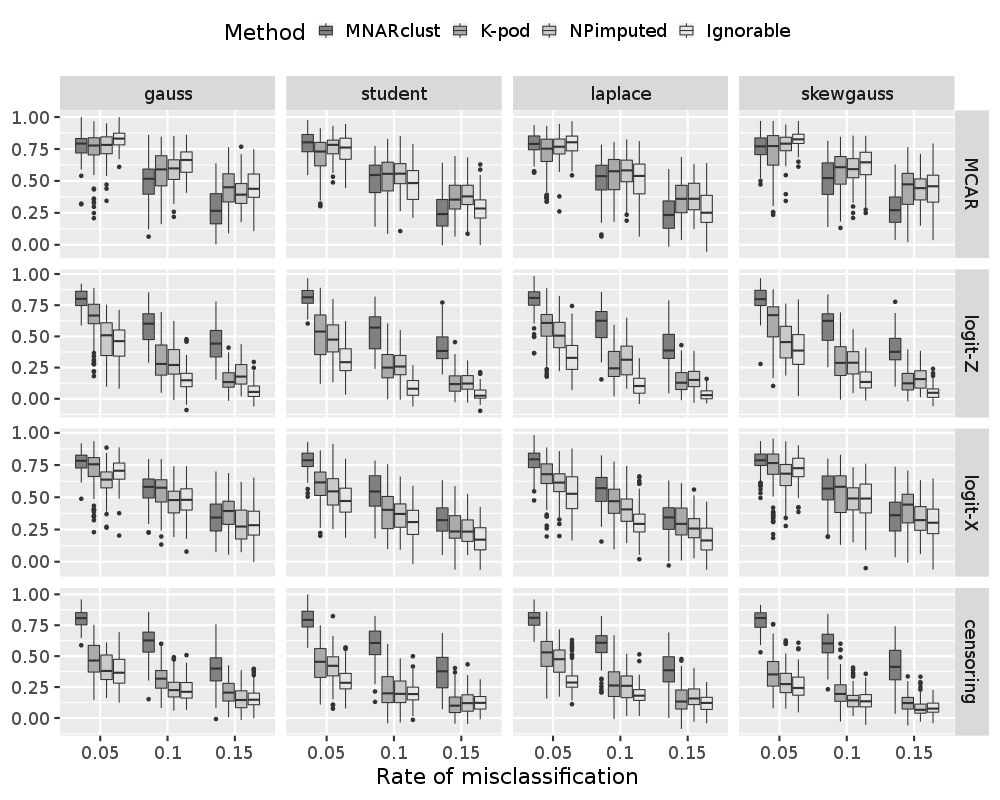}
\caption{Boxplot ARI obtained 100 samples composed of $n=100$ observations described by $d=6$ variables having a missing rate of $30\%$.} \label{fig:simu4}
\end{figure}

\paragraph{Impact of the theoretical misclassification}
To illustrate the impact of the overlaps between components, we consider data sets composed by $n=100$ observations generated with $d=6$ and a theoretical missing rate per variables of $30\%$ (parameters are given in \ref{tab:paramsim3}). For each scenario, we generated 100 data sets. Figure~\ref{fig:simu4} presents the boxplot of the ARI between the true partition and the estimators of the partition given by the methods. Overall, all the methods perform well under the MCAR mechanism despite that the results of the proposed methods are more deteriorated than those of the other methods when the misclassification rate is high. Howerver, under the non-ignorable scenarios, the proposed method outperforms the competing methods.

\subsection{Benchmark data}\label{subsec:appli1}
\begin{figure}[htp]
\centering \includegraphics[scale=0.35]{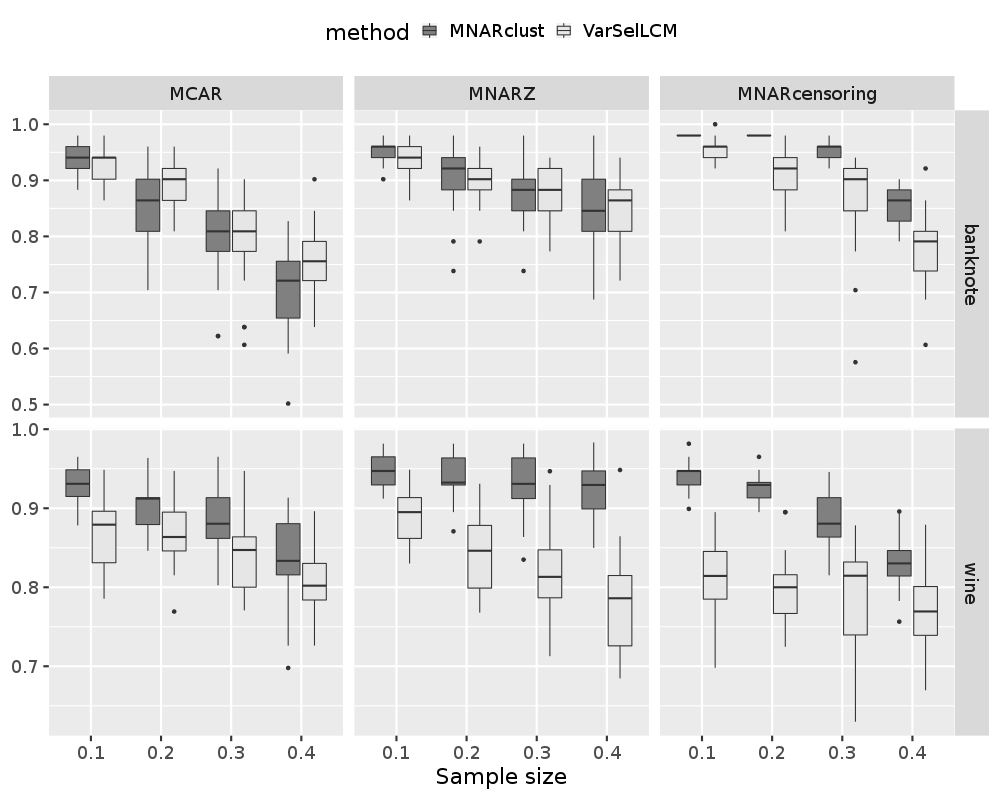}
\caption{Boxplot ARI obtained 25 samples composed genrated from the original data Swiss Banknotes (banknote) and Italian wines (wine).} \label{fig:bench}
\end{figure}

We consider two data sets (\emph{Swiss banknotes} and \emph{Italian wines}) described below to illustrate the behavior of the proposed method. The Swiss banknotes data set \citep{Flury1988} contains six measurements (length of bill, width of left edge, width of right edge, bottom margin width, top margin width and length of diagonal) made on 100 genuine and 100 counterfeit old-Swiss 1000-franc bank notes. This data set is available in the R package mclust \citep{ScruccaR2016}. The status of the banknote (genuine or counterfeit) is also known. We perform the clustering of the bills based on the six morphological measurements and we evaluate the resulting partition with the status of the bills. The Italian wine data set records 27 physical and chemical measurements on 178 Italian wines grown in the same region in Italy but derived from three different cultivars (Barbera, Barolo and Grignolino) and five years of production (1970, 1973, 1974, 1976 and 1979). The data set \citep{forinaVitis1986} is available on the R package MBCbook (companion R package of \citep{Bouveyron2019}). Clustering of the wines based on the 27 physical and chemical measurements and we compare the resulting partition with the three cultivars and the year of production. The orignial data does not have missing values. 
To investigate the behavior of the proposed method, we generate missing values on the original data using the following mecansims: \emph{MCAR} where the probability to unobserve each value is $\gamma$, \emph{MNARZ} where the probability to unobserve a value depends on the true class memberships (\emph{i.e.,} for the Swiss banknote and the Italian wine data sets, the true class memberships is defined by variable bill status and cultivars respectively; so this probability is $0.5\gamma$ and $\gamma$ for the counterfeit and genuine bills respectively and $0.5\gamma$, $\gamma$ and $1.5\gamma$ for the Barbera, Barolo and Grignolino respectively), \emph{MNARcensoring} where a variable is full observed if its value is more than the empirical quantile of the variable at level $\sqrt{\gamma}$ and observed with probability $\sqrt{\gamma}$ otherwise. Figure~\ref{fig:bench} presents the boxplot of the ARI obtained by considering a non-ignorable mechanism  of missingness (MNARclust) and an ignorable mechanism of missingness (VarSelLCM) on 25 samples generated under each scenario and for different values of $\gamma$. Note that on the original data, the semiparametric mixture demol and the Gaussian mixture model with conditional independence are relevant to detect the underlying partition. Indeed, on the Swiss banknote data set, the partition given by MNARclust and VarSelLCM has an ARI equal to 0.98 and 0.96. Moreover, on the Italian wine data set, the partition given by MNARclust and VarSelLCM has an ARI equal to 0.94 and 0.90. Results show that increasing the rate of missigness deteriorates the partitions. This phenomenon was expected because less discriminative information is present in the data set. Results show that the proposed method performs well under non-ignorable scenario while the results obtained by ignoring the mechanism are strongly deteriorated in such case (see MNARcensoring and MNARZ).

\subsection{Echocardiogram data set}\label{subsec:appli2}
We consider the \emph{Echocardiogram Data Set} \citep{salzberg1988exemplar} freely available in the R package \emph{MNARclust}.  This data set is composed by $n=132$ subjects who suffered from heart attack at some point in the past. The task is generally to determine from the other variables whether or not the patient will survive at least one year. The data set is composed by 5 continuous variables: \emph{age at heart attack} (missing rate $4.5\%$), \emph{fractional shortening} (a measure of contracility around the heart, lower numbers are increasingly abnormal, missing rate $6.0\%$), \emph{epss} (E-point septal separation, another measure of contractility, larger numbers are increasingly abnormal, missing rate $11.4\%$), \emph{lvdd} (left ventricular end-diastolic dimension; this is a measure of the size of the heart at end-diastole; large hearts tend to be sick hearts, missing rate $8.3\%$) and \emph{wall-motion-score} (a measure of how the segments of the left ventricle are moving, missing rate $3.0\%$); one binary variable \emph{pericardial effusion} (pericardial effusion is fluid around the heart, 0=no fluid, 1=fluid, missing rate $0.7\%$). We also have one binary variable which can be used as a partition among the subjects:  \emph{still alive} (0=dead at end of survival period, 1 means still alive). This binary variable is not used for clustering but permits to evaluate the accuracy of the estimated partition. 
Among the variables used for clustering there is 5.7\% of missing values and 19.1\% of the sujects have at least one missing value.  Moreover, the variable \emph{still alive} has only one missing value.
 
 Clustering is performed by extending the proposed approach to the case of mixed-type data (data set composed of one binary and five continuous variables). This extension, detailled in \ref{app:extension}, is easy because of the assumption of conditional independence within components. Hence, each categorical variable is modelled by a multinomial distribution given the component and the fact that the variable is observed. Moreover, since non-parametric estimation is only performed for the densities, smoothing is only done for the continuous variables.
 
Chosing the number of components in a semi-parametric mixture is a difficult problem (even in the complete case). Two recent methods \citet{kasaharaJRSSB2014, kwon2019estimation} have been developed to select this number in the case of continuous data set. Howerver, they cannot be used directly on mixed-type data. Thus, based on the evolution of the maximum smoothed log-likelihood with respect to the number of clusters (see Figure~\ref{fig:Kecho} in \ref{app:echo}), we chose to select $K=3$ clusters.  Figures~\ref{fig:echodiscrim1} and \ref{fig:echodiscrim2} presented in \ref{app:echo} show the relation between the minssingness rates and the influence on the missingess and on the observed variables on the partition. This quantity is measured by the empirical counterpart of $\mathbb{E}[\max_k \ln \mathbb{P}(Z_{ik}\mid R_{ij})]$ and $\mathbb{E}[\max_k \ln \mathbb{P}(Z_{ik}\mid X_{ij})]$ respectively. Thus, the more  these indexes are, there more discriminative the missigness process and the observed variable are. Moreover,  Figures~\ref{fig:echodiscrim1} and \ref{fig:echodiscrim2} show that both the missigness process and the observed variables influence the partition but that the observed variables are more discriminative (overall the values of $\mathbb{E}[\max_k \ln \mathbb{P}(Z_{ik}\mid R_{ij})]$ are less than thoses of $\mathbb{E}[\max_k \ln \mathbb{P}(Z_{ik}\mid X_{ij})]$).

 Table~\ref{tab:param} presents a summary of the conditional distribution of the variables given the clusters.
 \begin{table}[ht]
\centering
\begin{footnotesize}
\begin{tabular}{cccccccccccccccccc}
  \hline
 & \multicolumn{3}{c}{age } & \multicolumn{2}{c}{effusion} & \multicolumn{3}{c}{shortening} & \multicolumn{3}{c}{epss} & \multicolumn{3}{c}{lvdd} & \multicolumn{3}{c}{wall motion}\\ 
  \hline
  & $\tau_{kj} $ & mean & sd  & $\tau_{kj} $ & prob.  & $\tau_{kj} $ & mean & sd  & $\tau_{kj} $ & mean & sd  & $\tau_{kj} $ & mean & sd  & $\tau_{kj} $ & mean & sd\\
class-1 & 0.95 & 64.61 & 8.91 & 1.00 & 0.07 & 1.00 & 0.15 & 0.07 & 0.97 & 20.01 & 6.97 & 0.97 & 5.56 & 0.64 & 0.98 & 17.36 & 6.28 \\ 
  class-2 & 0.82 & 65.10 & 7.44 & 0.91 & 0.03 & 0.28 & 0.16 & 0.08 & 0.28 & 9.68 & 2.07 & 0.10 & 5.30 & 0.08 & 0.82 & 11.99 & 8.05 \\ 
  class-3 & 0.97 & 61.83 & 7.97 & 1.00 & 0.09 & 1.00 & 0.25 & 0.11 & 0.93 & 8.88 & 4.53 & 1.00 & 4.43 & 0.62 & 0.99 & 13.50 & 3.12 \\ 
   \hline
\end{tabular}
\end{footnotesize}
\caption{Summary of the conditional distribution of the vairables given the cluster: probability of non missing ($\tau_{kj}$), mean and standard deviation (sd) for the continuous variables and probability of occuring for the binary variable}\label{tab:param}
\end{table}

The three unbalanced classes are mainly explained by two variables: \emph{epss} and \emph{lvdd}, which are highly discriminative for both the missingness mechanism and the conditional densities $p_{kj}$. The four estimated classes can be described as follows:
 \begin{itemize}
 \item  \emph{class-1} ($\pi_1=0.27$) is composed of 33 subjects. These subjects are characterized by high values of the measures of \emph{epss}, \emph{lvdd} and  \emph{wall-motion-score} and small of values of the measures of \emph{fractional shortening}. This class is characterized by a very low probability of missingness for each variables;
 \item \emph{class-2} ($\pi_2=0.08$) is composed of 11 subjects. These subjects  have suffered from heart attack being older than the subjects of the other class and obtain low values for the \emph{wall-motion-score}. They are characterized by high probabilities of missingness for all the variables;
 \item \emph{class-3} ($\pi_3=0.65$) is composed of 88 subjects. These subjects  have suffered from heart attack being young and has low missingness probabilities. They take low values for \emph{epss} and \emph{lvdd} and high values of \emph{shortening}
\end{itemize} 
As shown by the confusion matrix presented in Table~\ref{tab:confalive}, the estimated partition permits to partially explain the death of the subject at the end of the survival period.

\begin{table}[ht!p]
\begin{center}
\begin{tabular}{cccc}
\hline
& Class 1 & Class 2 & Class 3  \\
\hline
dead&  12 & 4 & 72 \\
still alive & 21 & 6 & 16\\
\hline
\end{tabular}
\caption{Confusion matrix between the estimated partition and the variable \emph{still alive}. Adjusted Rand index is 0.25} \label{tab:confalive}
\end{center}
\end{table}

Finally, the assumption of independence within components seems to be realistic. Indeed, we investigate this assumption by testing the significance of the correlation coefficients between the conditional distribution of variables $X_{ij}$ given the cluster membership and $R_{ij}=1$. 
Table~\ref{pvalues2} and Table~\ref{pvalues3} in \ref{app:echo} presents the p-value obtained by testing the nullitiy of the correlation coefficient of the conditional distribution of couple of variables conditionnally on component 1 and 3 respectively. The high values of the p-values suggest that the assumption of conditional independence given the component membership is suitable. Note that results related to component 2 are not presented due a lack of subject affected to this class.

\section{Conclusion}\label{sec:concl}
The proposed method allows continuous data set with non-ignorable missingness to be clustered with no more assumption than the independence within components.
In some applications, the assumption of independence within components can be too strong but it can be relaxed. 
To consider dependencies, within components, of the missingness mechanism, the conditional distribution of $\bR_i\mid\bZ_i$ can be modeled by a dependence tree \citep{chow1968approximating}. This model considers, for each component $k$, a mapping $\sigma_k(j)$. Thus, we have
\begin{equation}\label{eq:tree}
 g_{k}(\br_i;\btau_k) = \prod_{j=1}^d\tau_{kj1}^{r_{ij}r_{i\sigma_k(j)}}(1-\tau_{kj1})^{(1-r_{ij})r_{i\sigma_k(j)}}\times\\\tau_{kj0}^{r_{ij}(1-r_{i\sigma_k(j)})}(1-\tau_{kj1})^{(1-r_{ij})(1-r_{i\sigma_k(j)})},
\end{equation}
where, by definition, there exists one $j_0\in\{1,\ldots,d\}$ such that $\sigma_k(j_0)=0$. We set $r_{i0}=1$ and $\tau_{kj_01}=\tau_{kj_00}$ which is the marginal probability of that variable $j_0$ to be observed under component $k$. Note that the marginal distribution of $\bR_i$ is a mixture of trees \citep{meilaJMLR2000}. This model is known to be flexible and easily interpretable. However, note that other distributions for multivariate binary data could be considered (see for instance \citet{weir2000binary,panagiotelisJASA2012,marbacCSDA2017}). Alternatively, blocks of within-components dependent variables can be considered to relax the within-component independence assumption (see \citet{levineBiometrika2011}, \citet{chauveauSurveys2015}). 

The approach could be extended to location or location/scale semi-parametric models. However, we believe that these models would be more suitable to model the distribution of the variables  than to model the conditional distribution of the variables given that their values are not missing.  However, location or location/scale semi-parametric models would permit to analyse univariate or bivariate data sets (that cannot be analyzed by our approach due to the assumption of Lemma~\ref{lem:iden}).

In the context without missingness, a drawback of the MM algorithm is the computation of integrals having no closed-form for computing the smoothed probabilities of subpopulation memberships. However, due to the independence within components, those integrals are only univariate. Note that the parametric mixtures (\emph{e.g.,} Gaussian mixtures) do not suffer from this drawback, when the data are complete. However, when missingness occurs, even the estimation of the parametric mixtures via EM algorithm leads to compute integrals having no closed form (see \cite{miaoJASA2016}). Thus, when missingness occurs, the estimation of the proposed semiparametric mixture is not more complex than the estimation of parametric mixture.

In this paper, we do not investigate kernel and bandwidth selection. Indeed, these selections are still an open question even  in the complete-data case. Thus, this problem deserves its own study that is out of the scope of this paper.  Finally, selecting the number of components is a difficult task for semiparametric mixture. Note that this task could be achieved by extending the approaches of \citet{kasaharaJRSSB2014} and \citet{kwon2019estimation} to mixed-type data.

\section*{Bibliography}

\bibliography{biblio}
\bibliographystyle{apalike}

\appendix
\section{Proofs}\label{appendix:proof}
\begin{proof}[Proof of Lemma~\ref{lem:iden}]
The model defined by \eqref{eq:melangeobs1}-\eqref{eq:melangeobs2} is identifiable, if
\begin{equation}\label{eq:ident}
\forall (\br_i^\top,\bx_i^\top)^\top \in \{0,1\}^d\times\mathbb{R}^d,\;
g(\bxo_i,\br_i;\btheta)=g(\bxo_i,\br_i;\tilde \btheta)
\;
\Rightarrow
\;
\btheta=\tilde \btheta,
\end{equation}
where $\btheta$ groups the finite dimensional parameters, $\pi_k$ and $\tau_{kj}$, and the infinite dimensional parameters $p_{kj}$ for $k=1,\ldots,K$ and $j=1,\ldots,d$. 
 Thus, considering the case where all the variables are observed (\emph{i.e,} $r_{ij}=1$, for $j=1,\ldots,d$),  the left hand side of \eqref{eq:ident} implies
\begin{equation}\label{eq:prodmixture}
\forall \bx_i\in\mathbb{R}^d,\; \sum_{k=1}^K \rho_k\prod_{j=1}^dp_{kj}(x_{ij})=\sum_{k=1}^K \tilde\rho_k\prod_{j=1}^d\tilde p_{kj}(x_{ij}),
\end{equation}
where $\rho_k=\pi_k\prod_{j=1}^d \tau_{kj}$ and $\tilde\rho_k=\tilde\pi_k\prod_{j=1}^d \tilde\tau_{kj}$.
Theorem~8 in \citet{allman2009identifiability} states that a mixture whose components are defined as product of univariate densities is identifiable if all the univariate densities are linearly independent and if $d\geq 3$, up to label swapping. Thus, under the conditions of Lemma~\ref{lem:iden}, Theorem~8 in \citet{allman2009identifiability} implies that $\forall k=1,\ldots,K,$ and $\forall j=1,\ldots,d,$
\begin{equation}\label{eq:step1}
 \pi_k\prod_{j=1}^d\tau_{kj}=\tilde \pi_k\prod_{j=1}^d\tilde\tau_{kj} \text{ and }  p_{kj}=\tilde p_{kj}.
\end{equation} 
This results and the left hand side of \eqref{eq:ident} imply
\begin{equation}\label{eq:step2}
\forall (\br_i^\top,\bx_i^\top)^\top \in \{0,1\}^d\times\mathbb{R}^d,\;
\sum_{k=1}^K \pi_k\prod_{j=1}^d\tau_{kj}\prod_{j=1}^d p_{kj}(x_{ij})
=
\sum_{k=1}^K \tilde \pi_k\prod_{j=1}^d\tilde\tau_{kj}\prod_{j=1}^d p_{kj}(x_{ij}).
\end{equation}
Considering the marginal distribution of $(r_{ij}^\top,x_{ij}^\top)^\top$ with $r_{ij}=1$, for any $j=1,\ldots,d$, we have from \eqref{eq:step2}
\begin{equation}\label{eq:step3}
\forall j=1,\ldots,d,\; \forall x_{ij}\in\mathbb{R},\;
\sum_{k=1}^K (\pi_k\tau_{kj} - \tilde \pi_k\tilde\tau_{kj})  p_{kj}(x_{ij})
=0.
\end{equation}
The densities $p_{kj}$ are linearly independent, so $\forall (\alpha_1,\ldots,\alpha_K)^\top\in\mathbb{R}^K\setminus\{\boldsymbol{0}\}$, $\sum_{k=1}^K \alpha_k p_{kj}$ is not the zero function. Thus, \eqref{eq:step3} implies that 
\begin{equation}\label{eq:step4}
\forall j=1,\ldots,d,\; \forall k=1,\ldots,K,\;  \pi_k\tau_{kj} = \tilde \pi_k\tilde\tau_{kj}.
\end{equation}
From \eqref{eq:step1} and \eqref{eq:step4}, recalling that $\pi_k>0$ and $\tau_{kj}>0$, for $k=1,\ldots,K$ and $j=1,\ldots,d$, we obtain that
$$
\forall k=1,\ldots,K,\;
\boldsymbol{M}\boldsymbol{u}_k=\boldsymbol{0},
$$
where
$$
\boldsymbol{M}=\begin{bmatrix}
1 & 1 & \ldots &1 \\
1 & && \\
\vdots & & \boldsymbol{I}_d & \\
1 & & & 
\end{bmatrix}
\text{ and }
\boldsymbol{u}_k= \begin{bmatrix}
\log(\pi_k/\tilde\pi_k)\\
\log(\tau_{k1}/\tilde\tau_{k1})\\
\vdots\\
\log(\tau_{kd}/\tilde\tau_{kd})\\
\end{bmatrix},
$$
where $\boldsymbol{I}_d$ is the identity matrix of size $d$. As $\boldsymbol{M}$ has full rank for $d\geq 2$, we deduce that $\boldsymbol{u}_k=\boldsymbol{0}$ and thus $\pi_k=\tilde \pi_k$ and $\tau_{kj}=\tilde \tau_{kj}$ for $k=1,\ldots,K$ and $j=1,\ldots,d$.
\end{proof}

\begin{proof}[Proof of Lemma~\ref{lem:monot}]
This proof is similar to the proof of Theorem~1 of \citep{levineBiometrika2011} and is only given for ease of reading.
We have 
$$
\ell_n(\btheta) - \ell_n(\btheta^{[r]}) \geq b^{[r]}(\btheta) - b^{[r]}(\btheta^{[r]}),
$$
where $b^{[r]}(\btheta)=\sum_{i=1}^n \sum_{k=1}^K t_{ik}(\btheta^{[r]})\ln \left(\pi_k\mathcal{N}g_k(\bxo_i,\br_i;\btheta)\right)$. Indeed, using the concavity of the logarithm,
\begin{align*}
\ell_n(\btheta) - \ell_n(\btheta^{[r]})  &= \sum_{i=1}^n \ln \left(\sum_{k=1}^K t_{ik}(\btheta^{[r]}) \frac{\pi_k\mathcal{N}g_k(\bxo_i,\br_i;\btheta)}{\pi_k^{[r]}\mathcal{N}g_k(\bxo_i,\br_i;\btheta^{[r]})}\right) \\
&\geq \sum_{i=1}^n \sum_{k=1}^K t_{ik}(\btheta^{[r]}) \ln \frac{\pi_k\mathcal{N}g_k(\bxo_i,\br_i;\btheta)}{\pi_k^{[r]}\mathcal{N}g_k(\bxo_i,\br_i;\btheta^{[r]})}\\
&= b^{[r]}(\btheta) - b^{[r]}(\btheta^{[r]}).
\end{align*}
To prove that the algorithm is monotone, it suffices to show that $\btheta^{[r+1]}$ is such that $b^{[r]}(\btheta) - b^{[r]}(\btheta^{[r]})\geq 0$. Note that the following decomposition holds
$$
b^{[r]}(\btheta) = b_1^{[r]}(\btheta) +b_2^{[r]}(\btheta) + \sum_{k=1}^K\sum_{j=1}^d b^{[r]}_{3kj}(\btheta)
$$
where $$b_1^{[r]}(\btheta) = \sum_{i=1}^n \sum_{k=1}^K t_{ik}(\btheta^{[r]})\ln\pi_k,\quad b_2^{[r]}(\btheta) = \sum_{i=1}^n \sum_{k=1}^K \sum_{j=1}^d t_{ik}(\btheta^{[r]}) \left(r_{ij}\ln\tau_{kj} + (1-r_{ij})\ln(1-\tau_{kj})\right),$$ and
  $$
  b_{3kj}^{[r]}(\btheta)=\int_{\Omega_j} \sum_{i=1}^n t_{ik}(\btheta^{[r]}) r_{ij} \frac{1}{h_j} K\left(\frac{x_{ij} - u}{h_j}\right)\ln p_{kj}(u)du.
    $$
Maximizing $b^{[r]}(\btheta)$ on the proportions $\pi_1,\ldots,\pi_k$ is equivalent to maximizing $b_1^{[r]}(\btheta)$ on the proportions. Similarly, maximizing $b^{[r]}(\btheta)$ on the probabilities $\tau_{kj}$ is equivalent to maximizing $b_2^{[r]}(\btheta) $ on the $\tau_{kj}$'s. Thus, one can check that the estimators $\pi_k^{[r+1]}$'s and $\tau_{kj}^{[r+1]}$'s maximize $b^{[r]}(\btheta)$ on the $\pi_k$'s and on the $\tau_{kj}$'s. Finally, note that we have
\begin{equation*}
b_{3kj}^{[r]}(\btheta) = - c_{kj}^{[r]} \int_{\Omega_j} p^{[r+1]}_{kj}(u) \ln \frac{p^{[r+1]}_{kj}(u)}{p_{kj}(u)} du + \\c_{kj}^{[r]} \int_{\Omega_j} p^{[r+1]}_{kj}(u) \ln  p^{[r+1]}_{kj}(u) du,
\end{equation*}
where $c_{kj}^{[r]}=\sum_{i=1}^n t_{ik}(\btheta^{[r]})r_{ij}$. The second term of the right hand side of the equation does not depend on $p_{kj}$. The first term of the right hand side of the equation is based on Kullback-Leibler divergence from $p_{kj}$ to $p^{[r+1]}_{kj}$. Thus, noting that $c_{kj}^{[r]}\geq 0$, $p^{[r+1]}_{kj}$ is the unique, up to changes on a set of Lebesgue measure zero, density function maximizing $b_{3kj}^{[r]}(\btheta)$. Proof is concluded by noting that $\btheta^{[r+1]}=\argmax_{\btheta}b^{[r]}(\btheta) $ leading that $ b^{[r]}(\btheta^{[r+1]}) \geq  b^{[r]}(\btheta^{[r]})$ and thus $\ell_n(\btheta^{[r+1]}) \geq \ell_n(\btheta^{[r]})$.
\end{proof}
%

\section{Simulation}\label{appendix:exp}
This section gives the values of $\delta$ and $\gamma$ used during the different experiments. These values have been estimated by generating a large sample ($n=10^4$ observations) where the misclassifaction rate is compute between the true partition and the partition given by the rule of the maximum \emph{a posteriori} computed with the true parameters. Note that, because the missingness mechanism impacts the distribution within components and thus the overlaps between components, changing the value of $\gamma$ permits to change the rate of missingness but implies changing the value of $\delta$ to keep hold the rate of misclassification. Table~\ref{tab:paramsim1} presents the parameters used  to generate the data with $K=3$ components, $d=6$ variables and a theoretical misclassification of $10\%$ (related to Figure~\ref{fig:simu1} and Figure~\ref{fig:simu2}).

\begin{table}[ht!p]
\centering
\begin{tabular}{cccccccccc}
  \hline
  Mechanism & Missing rate & \multicolumn{2}{c}{Gaussian} & \multicolumn{2}{c}{Student} & \multicolumn{2}{c}{Laplace} & \multicolumn{2}{c}{Skew-Gaussian}\\
  & & $\gamma$ & $\delta$& $\gamma$ & $\delta$& $\gamma$ & $\delta$& $\gamma$ & $\delta$ \\ 
  \hline
 MCAR & 0.200 & 1.380 & 1.817 & 1.391 & 2.437 & 1.388 & 2.351 & 1.409 & 1.304 \\ 
   & 0.300 & 0.837 & 2.013 & 0.849 & 2.717 & 0.838 & 2.609 & 0.866 & 1.420 \\ 
   & 0.400 & 0.402 & 2.310 & 0.400 & 3.220 & 0.403 & 3.032 & 0.416 & 1.625 \\ 
   logit-Z & 0.200 & -1.422 & 1.309 & -1.435 & 1.707 & -1.424 & 1.588 & -1.423 & 0.896 \\ 
   & 0.300 & -2.106 & 1.215 & -2.113 & 1.545 & -2.097 & 1.491 & -2.090 & 0.832 \\ 
   & 0.400 & -2.733 & 1.143 & -2.728 & 1.445 & -2.706 & 1.407 & -2.727 & 0.776 \\ 
   logit-X & 0.200 & 1.225 & 1.697 & 1.325 & 2.162 & 1.307 & 2.059 & 0.507 & 1.245 \\ 
   & 0.300 & 0.548 & 1.775 & 0.546 & 2.222 & 0.541 & 2.159 & -0.118 & 1.311 \\ 
   & 0.400 & -0.050 & 1.858 & -0.129 & 2.341 & -0.108 & 2.292 & -0.668 & 1.431 \\ 
   censoring & 0.200 & -0.549 & 1.594 & -0.645 & 2.060 & -0.587 & 1.962 & 0.313 & 1.101 \\ 
  & 0.300 & -0.181 & 1.598 & -0.205 & 2.064 & -0.180 & 1.957 & 0.553 & 1.114 \\ 
   & 0.400 & 0.159 & 1.611 & 0.171 & 2.065 & 0.118 & 1.949 & 0.795 & 1.142 \\ 
   \hline
\end{tabular}
\caption{Parameters used to generate the data with $K=3$ components, $d=6$ variables and a theoretical misclassification of $10\%$.} \label{tab:paramsim1}
\end{table}

 Table~\ref{tab:paramsim2} presents the parameters used  to generate the data with  $K=3$ components, a theoretical missing rate per variable of $30\%$ and a theoretical misclassification of $10\%$ (related to Figure~\ref{fig:simu3}). 
 
\begin{table}[ht!p]
\centering
\begin{tabular}{cccccccccc}
  \hline
  Mechanism & $d$ & \multicolumn{2}{c}{Gaussian} & \multicolumn{2}{c}{Student} & \multicolumn{2}{c}{Laplace} & \multicolumn{2}{c}{Skew-Gaussian}\\
  & & $\gamma$ & $\delta$& $\gamma$ & $\delta$& $\gamma$ & $\delta$& $\gamma$ & $\delta$ \\
  \hline
  MCAR & 3 & 0.831 & 3.340 & 0.847 & 5.566 & 0.849 & 4.850 & 0.834 & 2.384 \\ 
    & 6 & 0.837 & 2.013 & 0.849 & 2.717 & 0.838 & 2.609 & 0.866 & 1.420 \\ 
  & 9 & 0.847 & 1.611 & 0.853 & 2.099 & 0.839 & 1.997 & 0.846 & 1.132 \\ 
  logit-Z & 3 & -2.095 & 2.090 & -2.098 & 2.840 & -2.069 & 2.804 & -2.113 & 1.459 \\ 
    & 6 & -2.106 & 1.215 & -2.113 & 1.545 & -2.097 & 1.491 & -2.090 & 0.832 \\ 
  & 9 & -2.101 & 0.816 & -2.081 & -1.034 & -2.096 & 0.955 & -2.115 & 0.560 \\ 
  logit-X & 3 & 0.409 & 2.500 & 0.414 & 3.287 & 0.399 & 3.235 & -0.267 & 1.867 \\ 
 & 6 & 0.548 & 1.775 & 0.546 & 2.222 & 0.541 & 2.159 & -0.118 & 1.311 \\ 
  & 9 & 0.606 & 1.458 & 0.645 & 1.790 & 0.612 & 1.731 & -0.089 & 1.082 \\ 
    censoring & 3 & -0.146 & 2.243 & -0.161 & 3.095 & -0.140 & 2.972 & 0.577 & 1.596 \\ 
 & 6 & -0.181 & 1.598 & -0.205 & 2.064 & -0.180 & 1.957 & 0.553 & 1.114 \\ 
 & 9 & -0.207 & 1.327 & -0.242 & 1.650 & -0.209 & 1.556 & 0.537 & 0.916 \\ 
   \hline
\end{tabular}
\caption{Parameters used to generate the data with $K=3$ components, a theoretical missing rate per variable of $30\%$ and a theoretical misclassification of $10\%$.} \label{tab:paramsim2}
\end{table}

Table~\ref{tab:paramsim3} presents the parameters used to generate the data with $K=3$ components, $d=6$ variables and a theoretical missing rate per variable of $30\%$ (related to Figure~\ref{fig:simu4}).

\begin{table}[ht!p]
\centering
\begin{tabular}{cccccccccc}
  \hline
  Mechanism & $d$ & \multicolumn{2}{c}{Gaussian} & \multicolumn{2}{c}{Student} & \multicolumn{2}{c}{Laplace} & \multicolumn{2}{c}{Skew-Gaussian}\\
  & & $\gamma$ & $\delta$& $\gamma$ & $\delta$& $\gamma$ & $\delta$& $\gamma$ & $\delta$ \\
  \hline
  MCAR & 0.050 & 0.837 & 2.551 & 0.850 & 3.677 & 0.839 & 3.519 & 0.865 & 1.780 \\ 
  & 0.100 & 0.837 & 2.013 & 0.849 & 2.717 & 0.838 & 2.609 & 0.866 & 1.420 \\ 
   & 0.150 & 0.838 & 1.688 & 0.849 & 2.225 & 0.838 & 2.068 & 0.865 & 1.201 \\ 
    logit-Z & 0.050 & -2.106 & 1.733 & -2.113 & 2.288 & -2.097 & 2.174 & -2.089 & -1.166 \\ 
   & 0.100 & -2.106 & 1.215 & -2.113 & 1.545 & -2.097 & 1.491 & -2.090 & 0.832 \\ 
  & 0.150 & -2.106 & 0.868 & -2.113 & -1.092 & -2.097 & -0.988 & -2.089 & 0.571 \\ 
   logit-X & 0.050 & 0.485 & 2.138 & 0.469 & 2.770 & 0.461 & 2.724 & -0.176 & 1.593 \\ 
 & 0.100 & 0.548 & 1.775 & 0.546 & 2.222 & 0.541 & 2.159 & -0.118 & 1.311 \\ 
  & 0.150 & 0.602 & 1.503 & 0.609 & 1.876 & 0.615 & 1.788 & -0.072 & 1.122 \\ 
   censoring & 0.050 & -0.156 & 1.917 & -0.178 & 2.581 & -0.151 & 2.528 & 0.563 & 1.346 \\ 
     & 0.100 & -0.181 & 1.598 & -0.205 & 2.064 & -0.180 & 1.957 & 0.553 & 1.114 \\ 
 & 0.150 & -0.209 & 1.357 & -0.231 & 1.734 & -0.202 & 1.635 & 0.541 & 0.949 \\ 
   \hline
\end{tabular}
\caption{Parameters used to generate the data with $K=3$ components, $d=6$ variables and a theoretical missing rate per variable of $30\%$.} \label{tab:paramsim3}
\end{table}

\section{Extension of the approach to mixed-type data} \label{app:extension}
This sections considers that $\bX_i=(\bX_{i1}^\top,\ldots,\bX_{i d_c}^\top,X_{i d_c +1},\ldots,X_{id})^\top$ is a $d$-variate vector of mixed-type such that the first $d_c$ elements are categorical and the last $d-d_c$ elements are continuous. Each categorical variable $\bX_{ij}=(X_{ij1},\ldots, X_{ij m_j } )^\top$, with $1 \leq j \leq d_c$, has $m_j$ levels and $X_{ijh}=1$ if subject $i$ takes level $h$ for variable $j$ and $X_{ijh}=0$ otherwise. The definition of $\bR_i=(R_{i1},\ldots,R_{id})^\top$ is unchanged leading that  $R_{ij}=1$ if variable $j$ is observed for subject $i$ and $R_{ij}=0$ otherwise. Similarily to section~\ref{sec:model}, we consider that the couples $(X_{ij},R_{ij})^\top$ are conditionally independant given $\bZ_i$. Thus, the conditionnal distribution of $R_{ij}$ given $Z_{ij}=1$ is a Bernoulli distribution with parameter $\tau_{kj}$. The conditionnal distributions of a continuous variable (\emph{i.e.,} $X_{ij}$ with $d_c + 1 \leq j \leq d$) given $Z_{ij}=1$ and $R_{ij}=1$ and given $Z_{ij}=1$ and $R_{ij}=0$ are defined by the densities $p_{kj}$ and $q_{kj}$ respectively. Finally, the conditionnal distributions of a categorical variable (\emph{i.e.,} $X_{ij}$ with $d_c \leq j \leq d_c$) given $Z_{ij}=1$ and $R_{ij}=1$ and given $Z_{ij}=1$ and $R_{ij}=0$ are defined by two multinomial distributions. We denote $\beta_{kj}=(\beta_{kj1},\ldots,\beta_{kjm_j})^\top$ the vector defining the multinomial distribution of variable $j$ (with $1\leq j \leq d_c$) given $Z_{ik}=1$ and $R_{ij}=1$. Thus, $0<\beta_{kjh}$ is the probability that subject $i$ takes level $h$ for variable $j$ under component $k$ when this variable is observed and $\sum_{h=1}^{m_j} \beta_{kjh}=1$. Similarily to \eqref{eq:melangeobs1} and \eqref{eq:melangeobs2}, the distribution of the observed variables $(\bx_i^{\text{obs}\top},\br_i^\top)^\top$
\begin{equation*}
g(\bxo_i,\br_i;\btheta)=\sum_{k=1}^K \pi_kg_k(\bxo_i,\br_i;\btheta),
\end{equation*}
where  the pdf of component $k$ is a specific version of \eqref{eq:compo} defined by 
\begin{equation*}
 g_k(\bxo_i,\br_i;\btheta) = \left( \prod_{j=1}^d\tau_{kj}^{r_{ij}}(1-\tau_{kj})^{1-r_{ij}} \right) \left(\prod_{j=1}^{d_c} \prod_{h=1}^{m_j} \left(\beta_{kjh}\right)^{r_{ij} x_{ijh}}\right)\left(\prod_{j=d_c+1}^d p_{kj}^{r_{ij}}(x_{ij})\right).
\end{equation*}
Thus, a sufficient condition for identifiability, is that the proportions are not zero (\emph{i.e.,} $0<\pi_k$ for any $k$) and that there are at least three continuous variables (\emph{i.e.,} $d-d_c\geq 3$) that have linearly dependent densities $p_{kj}$ and non-zero probability of observing these variables under each components (\emph{i.e.,} $0<\tau_{kj}$).

The estimation is performed by maximizing the smoothed log-likelihood where the smoothing is performed only on the continuous variables. Therefore, the smoothed log-likelihood function is defined by
\begin{equation*}
\ell_n(\btheta) = \sum_{i=1}^n \ln\left( \sum_{k=1}^K \pi_k  \left( \prod_{j=1}^d\tau_{kj}^{r_{ij}} (1-\tau_{kj})^{1-r_{ij}} \right) \left(\prod_{j=1}^{d_c} \prod_{h=1}^{m_j} \left(\beta_{kjh}\right)^{r_{ij} x_{ijh}} \right) \left(\prod_{j=d_c+1}^d \exp\left(r_{ij} \mathcal{S} \ln p_{kj}(x_ij) \right) \right)\right).
\end{equation*}
In this context, the maximization on $\btheta$ of the smoothed log-likelihood function is performed via an MM algorithm. This iterative algorithm starts at the initial value of the parameters $\btheta^{[0]}$. At iteration $[r]$, it performs the following two steps
\begin{itemize}
\item Computing the smoothed probabilities of subpopulation memberships $$t_{ik}(\btheta^{[r]}) = \frac{\pi_k^{[r]} \mathcal{N}g_k(\bxo_i,\br_i;\btheta^{[r]})}{\sum_{\ell=1}^K\pi_\ell^{[r]} \mathcal N g_\ell(\bxo_i,\br_i;\btheta^{[r]})}.$$
\item Updating the estimators 
\begin{itemize}
\item Updating of the proportions
$$\pi_k^{[r+1]}= \frac{1}{n} \sum_{i=1}^n t_{ik}(\btheta^{[r]}).$$
\item Updating of the parameters of the missingness mechanism
$$ \tau_{kj}^{[r+1]} = \frac{\sum_{i=1}^n r_{ij} t_{ik}(\btheta^{[r]})}{\sum_{i=1}^n t_{ik}(\btheta^{[r]})}.$$
\item Updating of the parameters of the categorical variables, for $1\leq j \leq d_c$
$$ \beta_{kjh}^{[r+1]} = \frac{\sum_{i=1}^n r_{ij} x_{ijh} t_{ik}(\btheta^{[r]})}{\sum_{i=1}^n r_{ij} t_{ik}(\btheta^{[r]})}.$$
\item Updating of the conditional distribution of the continuous variables, for $d_c+1\leq j \leq d$
$$
p_{kj}^{[r+1]}(u) = \frac{\sum_{i=1}^n r_{ij} t_{ik}(\btheta^{[r]}) \frac{1}{h_j}K\left(\frac{x_{ij} - u}{h_j}\right)}{\sum_{i=1}^n r_{ij} t_{ik}(\btheta^{[r]})}.
$$
\end{itemize}
\end{itemize}

\section{Echocardiogram Data Set}\label{app:echo}
\begin{figure}[ht!]
\centering \includegraphics[scale=0.3 ]{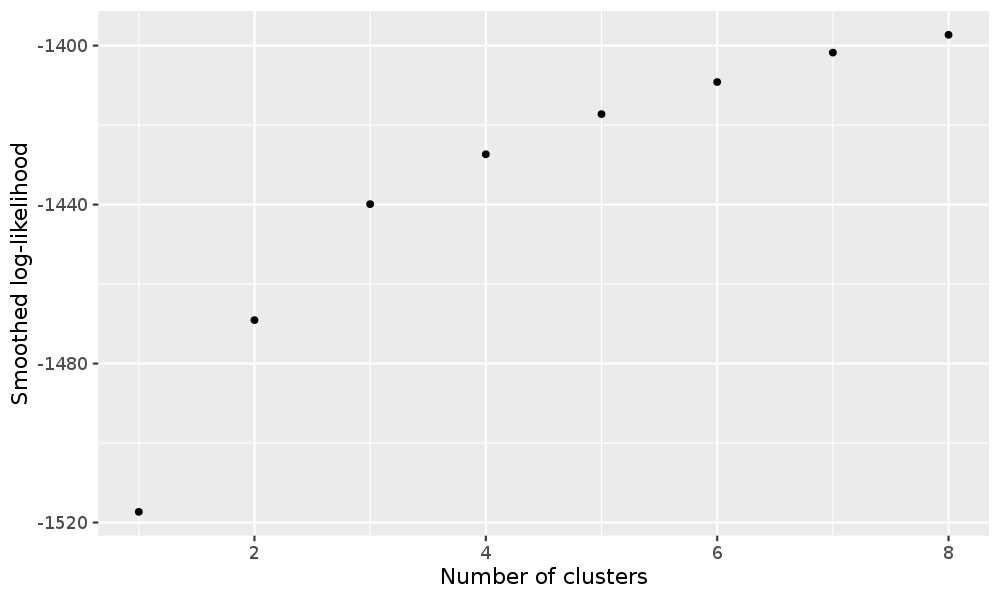}
\caption{Maximum of the smoothed log-likelihood with respect to the number of clusters}\label{fig:Kecho}
\end{figure}

Figure~\ref{fig:Kecho} helps to selected a suitable number of components by presenting the evolution of the maximum smoothed log-likelihood with respect to the number of clusters

\begin{figure}[ht!]
\centering \includegraphics[scale=0.3 ]{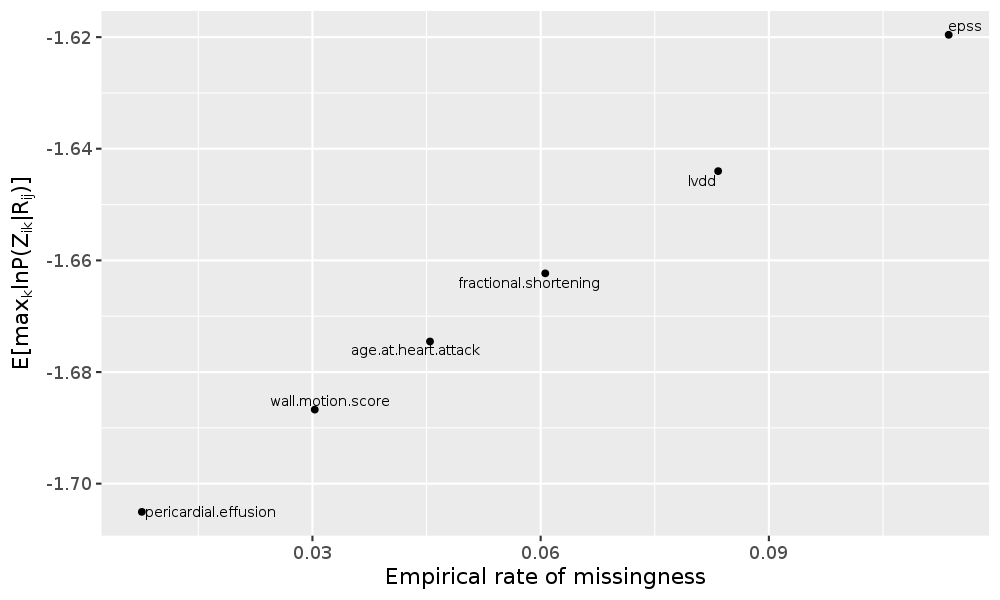}
\caption{Rate of missingness and empirical counterpart of $\mathbb{E}[\max_k \ln \mathbb{P}(Z_{ik}\mid R_{ij})]$ for each variable.}\label{fig:echodiscrim1}
\end{figure}

\begin{figure}[ht!]
\centering \includegraphics[scale=0.3 ]{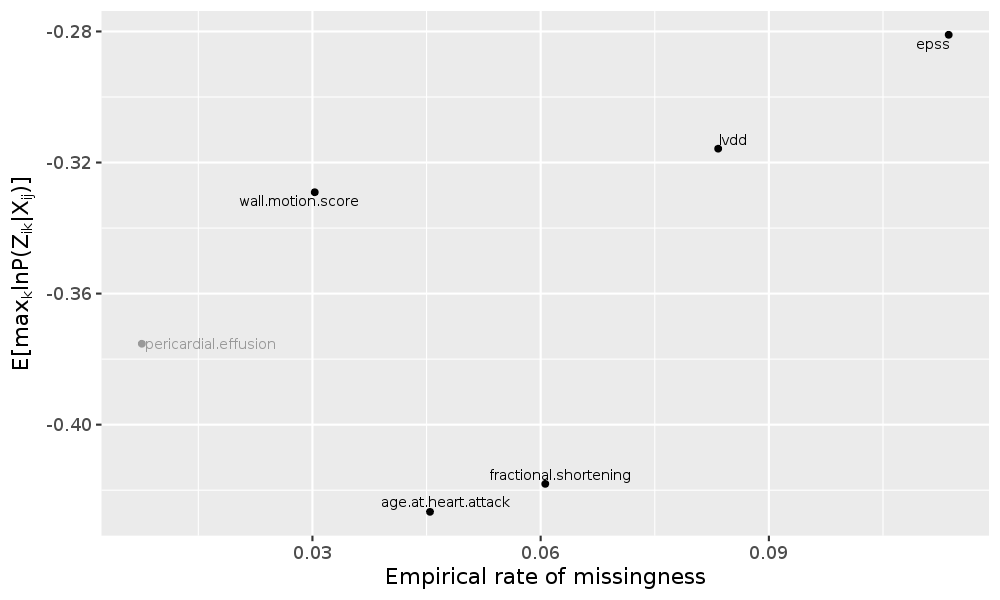}
\caption{Rate of missingness and empirical counterpart of $\mathbb{E}[\max_k \ln \mathbb{P}(Z_{ik}\mid X_{ij})$ for each variable.}\label{fig:echodiscrim2}
\end{figure}

Figure~\ref{fig:echodiscrim1} illustrates the relation between the rate of missingness and how the missingness of a variable is informative for the partition. Moreover,  Figure~\ref{fig:echodiscrim2}  illustrates the relation between the rate of missingness and how the observed values of a variable is informative for the partition.

Table~\ref{pvalues2} and Table~\ref{pvalues3} presents the p-value obtained by testing the nullitiy of the correlation coefficient of the conditional distribution of couple of variables conditionnally on component 1 and 3 respectively. The high values of the p-values suggest that the assumption of conditional independence given the component membership is suitable. Note that results related to component 2 are not presented due a lack of subject affected to this class.
\begin{table}[ht]
\centering
\begin{tabular}{cccccc}
  \hline
&age-at-attack&fractional-shortening&epss&lvdd&wall-motion  \\ 
  \hline
age-at-attack & 0.00 & 0.11 & 0.21 & 0.77 & 0.41  \\ 
fractional-shortening & 0.11 & 0.00 & 0.87 & 0.52 & 0.32\\ 
epss & 0.21 & 0.87 & 0.00 & 0.01 & 0.58 \\ 
lvdd & 0.77 & 0.52 & 0.01 & 0.00 & 0.48 \\ 
wall-motion & 0.41 & 0.32 & 0.58 & 0.48 & 0.00\\ 
   \hline
\end{tabular}
\caption{Pvalues obtained for testing the significance of the correlation coefficient computed on the couple of observed variables for observations affected to cluster 1.\label{pvalues2}}
\end{table}

\begin{table}[ht]
\centering
\begin{tabular}{cccccc}
  \hline
&age-at-attack&fractional-shortening&epss&lvdd&wall-motion  \\ 
  \hline
age-at-attack &0.00 & 0.50 & 0.84 & 0.22 & 0.65\\ 
fractional-shortening &0.50 & 0.00 & 0.06 & 0.17 & 0.80  \\ 
epss & 0.84 & 0.06 & 0.00 & 0.10 & 0.24 \\ 
lvdd & 0.22 & 0.17 & 0.10 & 0.00 & 0.03 \\ 
wall-motion & 0.65 & 0.80 & 0.24 & 0.03 & 0.00\\ 
   \hline
\end{tabular}
\caption{Pvalues obtained for testing the significance of the correlation coefficient computed on the couple of observed variables for observations affected to cluster 3.\label{pvalues3}}
\end{table}

\end{document}